\documentclass[Crown,sageh,times]{sagej}

\usepackage{moreverb,url}
\usepackage{tabularx}
\usepackage{booktabs}
\usepackage{amsmath}
\usepackage{graphicx}
\usepackage[colorinlistoftodos]{todonotes}
\usepackage{float}
\usepackage{subcaption}
\usepackage[export]{adjustbox}
\usepackage{setspace}
\usepackage{lineno}
\usepackage{dirtytalk}

\usepackage[colorlinks,bookmarksopen,bookmarksnumbered,allcolors=blue]{hyperref}

\newcommand\BibTeX{{\rmfamily B\kern-.05em \textsc{i\kern-.025em b}\kern-.08em
T\kern-.1667em\lower.7ex\hbox{E}\kern-.125emX}}

\begin{document}

\runninghead{Sanford and Yasseri}

\title{\Large The Kaleidoscope of Privacy:\\ Differences across French, German, UK, and US GDPR Media Discourse}
\vspace{-1cm}
\author{ \vspace{0cm}Mary Sanford\affilnum{1} and Taha Yasseri\affilnum{1,2,3,4}}

\affiliation{\affilnum{1}Oxford Internet Institute, University of Oxford, Oxford, UK\\
\affilnum{2}School of Sociology, University College Dublin, Dublin, Ireland\\
\affilnum{3}Geary Institute for Public Policy, University College Dublin, Dublin, Ireland\\
\affilnum{4}Alan Turing Institute, London, UK}

\corrauth{Taha Yasseri,
School of Sociology, University College Dublin, Belfield, D04 V1W8 Dublin, Ireland.\\ +353-1-7168561}
\email{taha.yasseri@ucd.ie}

\begin{abstract}
Conceptions of privacy differ by culture. In the Internet age, digital tools continuously challenge the way users, technologists, and governments define, value, and protect privacy. National and supranational entities attempt to regulate privacy and protect data managed online. The European Union passed the General Data Protection Regulation (GDPR), which took effect on 25 May 2018. The research presented here draws on two years of media reporting on GDPR from French, German, UK, and US sources. We use the unsupervised machine learning method of topic modelling to compare the thematic structure of the news articles across time and geographic regions. Our work emphasises the relevance of regional differences regarding valuations of privacy and potential obstacles to the implementation of unilateral data protection regulation such as GDPR. We find that the topics and trends over time in GDPR media coverage of the four countries reflect the differences found across their traditional privacy cultures.
\end{abstract}


\maketitle

The digital revolution in our societies and particularly the Internet-based technologies have complicated how individual users, business owners, and governments handle complex issues such as privacy. The ways in which many people use these technologies generate copious amounts of data that require varying degrees of protection. As such, governments around the world try to regulate how companies manage user data and establish standards of accountability. However, many of these attempts impose unilateral requirements on diverse contexts that may not have compatible value, legal, or political systems. The European Union's General Data Protection Regulation (GDPR) provides an example of such legislation that affects billions of people from a wide range of countries and cultures, each with unique values, political systems, and legal codes. 

The EU Parliament ratified GDPR on 24 May 2016 and it came into effect 25 May 2018. It replaces the 1995 European Data Protection Directive (DPD). It consolidates various country-specific data protection laws developed in addition to the DPD over the past two decades into a single standard. It establishes a cohesive approach to protecting privacy, increasing transparency of data collection and processing, and re-emphasising values of trust and accountability in the data economy \citep{uk_information_comissioners_office_gdpr_2017}. GDPR explicitly addresses issues such as data portability, consent to data collection and processing, children’s data, the right to be forgotten, and the general rights of private EU citizens to their online data profiles. Critically, GDPR affects any enterprise that handles the data of EU citizens, regardless of the company's provenance. 

Most of the academic literature related to GDPR focuses on its implications for specific industries, such as administrative research \citep{mourby_are_2018}, healthcare and medicine \citep{mccall_what_2018}, business \citep{tankard_what_2016}, and technology \citep{schweighofer_gdpr_2017}. Other research introduces tools or conceptual frameworks to guide businesses about GDPR compliance, e.g. privacyTracker \citep{casteleyn_privacytracker:_2016}, while still others explore how the regulations impacts privacy norms and ongoing challenges \citep{wachter_normative_2018}. In the years since its implementation, researchers have sought to understand the effects of GDPR. \cite{zaeem_effect_2020} analyse how GDPR has affected privacy policies around the globe, \cite{li_impact_2019} examine the effect of the legislation on global technology development, \cite{van_ooijen_does_2019} study the extent to which GDPR strengthens consumer control over their data, and \cite{aridor_economic_2020} analyse various empirical evidence of its economic impact. Yet, research on the public discourse of GDPR as reflected in the national and local media in different countries remains neglected. The research presented here seeks to fill this gap by providing a temporal analysis of the comparative emergence and clustering of themes in GDPR media discourse in order to investigate how this discourse reflects the diffusion of privacy traditions across four unique cultural settings.
We explore the following research questions: 

\begin{enumerate}
\item What trends exist in public media reporting on GDPR in the years leading up to its implementation across geographic region and source political leaning? 

\item Do these trends match the traditional conceptual diffusion of privacy across cultures? 

\item What logistic issues might we be able to forecast about the implementation of GDPR given any evident divergence of the discourse pertaining to it?

\end{enumerate}

To answer these questions we focus on news articles published in the mainstream media in the countries of interest. We focus on news articles because of the demonstrated role of public media in reflecting and shaping general opinion \citep{habermas}. However, we acknowledge that the public media is not the only place in which public sentiment related to GDPR discussed, and thus also influenced. For example, we do not study GDPR-related content shared on social media platforms. Nonetheless, as the digital media becomes more popular as a source of information and public opinion influence, it is important to study how cultural differences regarding values such as privacy manifest themselves in public media coverage of prominent regulation. Moreover, the study of media discourse provides insight to the sentiment, opinions, and values of a society. 

To analyse the articles, we use topic modelling: an unsupervised machine learning paradigm, which relies on Bayesian probability to determine the underlying structure amongst topics, also known as latent themes, in a corpus of text input \citep{lafferty_david}. Critically, topic models do not require prior labelling of the documents to be classified; the thematic structure emerges from the generative topic model \citep{blei_probabilistic_2012}. Topic modelling has been successfully used throughout social science research (examples given in the Methods section). 

\section*{An umbrella over a kaleidoscope} \label{back}
The concept of privacy and the extent to which EU lawmakers believe users are entitled to it lies at the centre of this legislation and the controversy surrounding it. Over the last half-century, dozens of scholars have published findings about the lack of a universally-accepted definition of privacy and the role of macro and individual-level cultural elements that influence the way a given culture defines and values privacy \citep{hichang_cho_multinational_2009}. Whitman describes privacy as \say{an unusually slippery concept $[~\cdots~]$ the sense of what must be kept \say{private}, or what must be hidden before the eyes of others, seems to differ strangely from society to society} \citep{whitman_two_2004}. As such, many scholars call for lawmakers to recognise these differences and avoid embedding the assumption of a universal privacy conception in the legislation they generate \citep{bellman_international_2004}. GDPR has the potential to do exactly what these scholars warn against. That is, it establishes data protection laws that are inherently  privacy-related, without accounting for the variation in how different cultures within the legislation's scope of impact value privacy. This failure has the potential to hinder the the legislation's implementation  and its effectiveness.   

Several studies have found continental European countries such as France and Germany, to have considerably different privacy values than those of the UK and US \citep{charlesworth_law_2003,merchant_privacy_2016,milberg_values_1995,milberg,post_three,schwartz_reconciling_2014,trouille_private_2000}. Furthermore, many of the companies impacted by GDPR, such as Facebook and Google, are companies incorporated in the US. In the years preceding GDPR, these companies fought EU governments repeatedly on matters of jurisdiction and accountability for their operations in the EU, those concerning EU citizens, or simply their data. For example, European courts have taken Goolge through lengthy judicial processes pertaining to the right to be forgotten \citep{delete,frantziou}. These cases resulted in Google changing its policies for certain subsets of its global user base.

\section*{Differences engraved in the law} 
Continental European, British and American cultures share several values, such as freedom of religion, opinion, and expression, as well as the inherent value of each individual \citep{merchant_privacy_2016}. Despite these similarities, privacy values in France, Germany, the UK, and the US vary significantly for several reasons, the most influential of which pertains to the source of the intrinsic meaning that the four countries give to privacy: the legal systems of the UK and US protect personal privacy to the extent that these efforts compliment and enable commercial endeavours; whereas France and Germany view privacy as a fundamental human right \citep{charlesworth_law_2003,schwartz_reconciling_2014}. 

The French and German constitutions include explicit provisions for individual privacy protection. The US Constitution does not have these provisions but it has been argued that in order for many of the other freedoms proclaimed in the Constitution to have value, it must also provide an \say{implicit right to privacy} \citep{charlesworth_law_2003}. Thus, the legal basis for general privacy standards in the US is implicitly derivable from the Constitution and thereafter arbitrated ad hoc. The UK, lacking a formal written constitution, resorts to similarly narrow or \say{sectoralized} parliamentary laws and international political and economic pressure for the development and maintenance of privacy standards \citep{charlesworth_law_2003}. In comparison with continental Europe, and to some extents the US, the UK traditionally demonstrates a much lower inclination to support and develop formal legal provisions for the protection of personal privacy \citep{charlesworth_law_2003}. 

\section*{Measuring the divide} 

\begin{figure}[ht]
\centering
\includegraphics[width=0.9\textwidth]{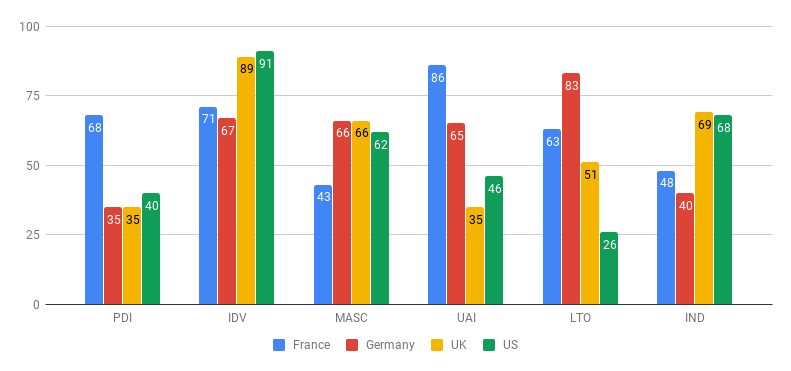}
\caption{\label{fig:hof}Scores for each of Hofstede's six cultural comparison dimensions for all countries. PDI: Power equality distance, IDV: Individualism versus collectivism, MASC: Masculinity versus femininity, UAI: Uncertainty avoidance, LTO: Long term orientation, IND:Indulgence.}
\end{figure}

To address the issue of cultural differences more systematically, we use Hofstede`s (2001) six-dimensional model of cultural comparison \citep{hofstede_cultures_2001}. Based on thorough research of individual countries around the world, Hofstede and his team score each country on a scale from 0 to 100 for the following dimensions: power equality distance (PDI), individualism versus collectivism (IDV), masculinity vs femininity (MASC), uncertainty avoidance (UAI), long-term versus short-term normative orientation (LTO), and indulgence versus restraint (IND). PDI measures the power equality in a culture, i.e.  power differential between the most and least powerful individuals in a society. IDV measures how much individuals in a society value loose social networks in which they care for themselves and their immediate families, versus tighter networks with more than one family or unit in which all members of the \say{in-group} are responsible for each other. MASC measures the extent to which individuals value material success, competition, and assertiveness versus consensus, quality of life, and cooperation. UAI measures how much uncertainty members of a society can tolerate. LTO measures how much the past history of a culture impacts how it shapes its present and future. Finally, IND measures the proclivity of individuals in a society towards unrestricted gratification of pleasure and enjoyment. Table \ref{tab:dims} summarises the dimensions and their definitions and Figure \ref{fig:hof} shows the distribution across each of the dimensions for France, Germany, the UK, and the US.

\begin{table}[ht]
\caption{Summary of Hofstede`s dimensions and their abbreviations \citep{hofstede_cultures_2001}.}
\label{tab:dims}
\begin{tabularx}{\textwidth}{ c p{0.33\textwidth} c c }
\toprule
Name  & Definition & Abbreviation \\
\midrule
Power equality distance & Difference in how much power the most and least powerful individuals in a society & PDI \\
Individualism versus collectivism & How much a society values looser versus tighter social networks  & IDV \\
Masculinity versus femininity & Extent to which individuals value material success, competition, and assertiveness & MASC \\
Uncertainty avoidance & How much uncertainty members of a society can tolerate in their environment & UAI \\
Long term orientation & How much the members of a society allow its past to inform its present and future & LTO \\
Indulgence & The tendency of individuals towards instant gratification & IND \\
\bottomrule
\end{tabularx}
\end{table}

Previous cultural studies of privacy identify the PDI, IVD, MASC, and UAI indices as the most relevant to cultural valuations of privacy \citep{milberg_values_1995,hichang_cho_multinational_2009}. \cite{milberg} draw the following conclusions about the correlations between these indices and privacy values and behaviours: cultures with high PDI scores tolerate more power inequality but also harbour greater mistrust of powerful groups; low IND scores correlate with greater acceptance of intrusion in one’s private life; cultures with high MASC scores tend to value material success more and are thus more likely to accept the benefits of using private information for economic purposes over the cost of intrusion; and finally, cultures with high UAI tend to prefer clear rules to reduce uncertainty and are more accepting of higher levels of privacy regulation. 

\section*{Methods} \label{datamethods}
We use LexisNexis to collect the articles. LexisNexis archives news articles from a wide range sources for dozens of countries and languages. We collected articles for each of the countries of interest using key terms \say{GDPR}, \say{RGPD}, and \say{DSGVO}. As referenced earlier, GDPR is the acronym for the legislation in English. RGPD is the acronym for the French name of the legislation. \say{DSGVO} is the acronym for the German name. We then merged the results of our searches together in order to eliminate duplicate articles and manually evaluated the metadata of each article to ensure relevance. LexisNexis allows you to search all German-language and French-language media in its archives, which inevitably includes sources from other French or German speaking countries that are not France or Germany. Thus, we manually removed sources if they fit this latter criteria.

All articles were published during the period of May 2016 to early June 2018. 
Our data collection period is determined by two major events: the regulation entered into force on 24 May 2016, and its provisions became directly applicable in all member states on 25 May 2018. For each article, we collected the publication date, source, title, and author. We collected 3274 articles in total: 325 from US media, 362 from UK media, 1242 from French media, and 1542 from German media.  
The distribution of articles is not uniform across the four languages. To determine whether or not this irregularity results from limitations of the scope of LexisNexis we examined the Google Trends results for our search terms: \say{GDPR}, \say{RGPD}, and \say{DSGVO.} The results suggest that the French and German media have reported much more frequently on GDPR than the UK and US, which is in accordance with the number of articles in each language in our collected corpus.

\subsection*{Data Processing} \label{dataproc}
To prepare the data for topic modelling, we removed all numbers, white spaces, punctuation, and lowercase all letters, and all stop words with the {\it tm} R package and manual inspection. 
Next, we removed any non-word terms and custom non-words, such as URLs, \say{etc}, \say{ie}, and \say{eg} that appear in the texts.
Additionally, we removed modal verbs such as \say{like}, \say{come}, \say{would}, \say{might}, etc., for the same reason as to why we remove the stop words.
Next, we remove words that occur very infrequently across all articles, i.e. words with high sparsity values, and words of substance that occur in virtually every document and can thus be ignored because as a result of their high frequency, they add noise to the topic distributions. The code we used to implement these steps is available at xxx.

\subsection*{Topic Modelling} \label{tm}
The most commonly used algorithm for topic modelling has been Latent Dirichlet Allocation (LDA), which we use here. LDA, as applied to topic modelling specifically, assumes a representation of each document as a distribution over latent topics (or themes) and that the topic distribution of all documents share a common prior drawn from a Dirichlet distribution. Additionally, it represents each topic as a distribution over words that also share a common Dirichlet prior. It uses this information to reconstruct the units of input after identifying their latent thematic structures. 
Additionally, LDA assumes that all texts or sources of content in a collection share the same set of topics but in different proportions relative to the others. The algorithm then seeks to determine the thematic or topical structure structure that likely generated the observed collection of texts. It bears the name \say{latent} Dirichlet allocation because the topic structure generated for each document in the corpus constitutes the \say{hidden structure} of the visible texts.  
In relation to the domain of the current work, topic modelling has been used for analysing the press successfully to study portrayals of Muslim women in The New York Times and The Washington Post over a thirty-five year period \citep{terman_islamophobia_2017}, to investigate the \say{frames} construed in the media as consequence of the controversies related to U.S. federal funding of the arts \citep{DIMAGGIO2013570}, and to compare the research output of the English and French demographic communities over a fifty-nine year period \citep{MARSHALL2013701}. 
In a more general context, topic modelling has been used to analyse self-reported accounts of sexist behaviour \citep{melville2019}, content of the petitions submitted to the UK Government petitioning website \citep{vidgen2020}, and the social media response to the 2019 IPCC Report on Climate Change and Land \citep{sanford2021}, among many other examples.

We train the model on each data set separately and find the hyperparameters\endnote{LDA topic modelling utilises three hyperparameters to calculate the distributions of words across topics and topics across documents in a corpus: $\alpha$, $\beta$, and $k$. $\alpha$ determines the ratio of documents to topics, $\beta$ determines the ratio of topics to words, and $k$ is the number of topics. We tuned these three hyperparameter by calculating the perplexity distribution across a range of possible values. For each potential value of a hyperparameter, we divided the data into five subsets, trained the model on four and tested on the fifth iteratively until all combinations of the five subsets have been tested. For $k$, we tested values 5-25 and use Kaiser’s elbow rule to determine the optimal value \citep{kaiser}.}
and the number of topics that lead to highest coherence\endnote{The coherence of each topic refers to how well, or how sensibly from a human perspective, the terms relate to one another semantically. Higher coherence scores indicate more cohesive topics and higher interpretive validity. The specific coherence score reported here is \cite{mimno_optimizing_nodate}'s UMass coherence, which is based on the document co-occurrence of individual words. The closer the values to 0 the more coherent the topics are} for each country data set (see Table \ref{tab:params}). We generate parallel sets of models, one with stemming and one without stemming,\endnote{Stemming is a way to reduce the size of a vocabulary by conflating words with related meanings \citep{mimno_optimizing_nodate}} and subsequently focus on the one with the highest cohesion and interpretability for each language. After running the topic modelling, we assign each topic a descriptive title based on the top fifty most frequent terms assigned to it by the model.

\begin{table}[ht]
\centering
\caption{\label{tab:parameters}Optimal hyperparameter values and the number of topics for each model.}
\label{tab:params}
\begin{tabular}{ c c c c }
\toprule
Model & $k$ & $\alpha$ & $\beta$ \\
\midrule
France-unstemmed & 9 & 0.2 & 0.1 \\
Germany-unstemmed & 12 & 0.4 & 0.1 \\
UK-stemmed & 6 & 0.1 & 0.5 5\\
US-stemmed & 7 & 0.3 & 0.1 \\
\bottomrule
\end{tabular}
\end{table}

\subsection*{Network Analysis}
To build a network of topics, we use cosine similarity to quantify the connection between the topics similar to \cite{melville2019} and \cite{vidgen2020}. Cosine similarity measures the distance between two non-zero vectors. In our case, each vector corresponds to a topic with as many elements as articles in the corpus. An element takes on the value 1 if the corresponding article contains the given topic and 0 if it does not. Higher similarity values between two topics means that those two topics appear frequently in the same articles. 

We ran the Louvain community detection algorithm \citep{blondel2008fast} to quantify the extent of topic clustering in each model. Clusters or communities in a topic network are identified by a greater amount of internal connections relative to the number of connections to topics outside the cluster. Community detection algorithms quantify this ratio for every possible community in a network and iteratively searches for the community allocation that maximises this ratio. The presence of distinct topic communities in our models could indicate distinct narratives or foci in the media discourse. 

\section*{Results and Discussion} \label{results}
The French and German datasets exhibit the largest range of sources, from media giants to regional and local outlets. This suggests that GDPR receives media attention at all levels of media reporting. The UK and US datasets are much smaller than the French and German ones, both in number of articles and variety of outlets. Additionally, the sources with the most representation in the US dataset are not the typical, well-established outlets. This could be the result of LexisNexis having lower access to US media sources due to pay restrictions, data protection issues, copyright complications, etc. However, we suggest that the lower rate of reporting on GDPR in the UK and US is not entirely an artefact of the LexisNexis database, but one that likely stems from less interest in the legislation as a result of GDPR not directly applying to the UK and US (in light of the EU referendum and the tradition of the UK to adapt EU law to its own system) and looser privacy values in these cultures. In the following we present the results for each country's data set.

\subsection*{France} \label{resfr}
Table \ref{tab:frtopics} shows the results of the French topic model. The model contains topics pertaining to the following subjects: the Facebook-Cambridge Analytica scandal, general data-related scandals, GDPR scope outside the EU, consumer rights and protection, social media and other data gathering applications, the role or involvement of the technology sector, Internet concerns, data security, and US - UK adoption of or cooperation with GDPR. Three of the topics produced by the model consist of difficult to annotate or miscellaneous terms and as such, we omit them here. The most coherent topics include the consumer rights and protection topic, the GDPR scope topic, the US-UK adoption topic, and the general data scandal topic. Interestingly, both the consumer rights and protection topic and the GDPR scope topic were also among the most prevalent topics. The prevalence analysis ranked the following topics highest in the model: GDPR scope, security, and consumer rights and protection. 

\begin{table}[ht]
\centering
\caption{\label{tab:frtopics}French model topic results with top 10 terms per topic and the prevalence (coherence score in parentheses) of each topic. FB/CA refers to Facebook and Cambridge Analytica.}
\begin{tabular}{@{\extracolsep{5pt}} ccccc}
\toprule
FB/CA & Internet & Tech Sector & Scope & Security \\
\midrule
\textit{45 (-50)} & \textit{67} (-75) & \textit{98 (-62)} & \textit{172 (-33)} & \textit{178 (-37)} \\
\midrule
facebook & personnelles & entreprises & donn\'es & s\'ecurit\'e \\
zuckerberg & droit & euros & num\'erique & clients \\
donn\'ees & loi & recherche & europe & gestion \\
parlement & nouveau & intelligence & entreprises & services \\ 
partenaire & compte & artificielle & vie & place \\
europ\'een & protection & innovation & google & donn\'ees \\
groupe & texte & num\'erique & priv\'ee & client \\
cambridge & site & pr\'esident & europ\'eenne & entreprise \\
analytica & service	& secteur & personnelles & temps \\
europ\'eens & utilisateur & tech & europ\'een & depuis	\\
\end{tabular}
\centering
\begin{tabular}{@{\extracolsep{5pt}} cccc}
\toprule
Social Media \& Apps & Scandal & Consumer Rights & US/UK \\
\midrule
\textit{48 (-46)} & \textit{64 (-28)} & \textit{199 (-26)} & \textit{75 (-27)} \\ 
\midrule
facebook & facebook & donn\'ees & facebook \\
donn\'ees & zuckerberg & protection & zuckerberg \\
google & donn\'ees & personnelles & mark \\
utilisateurs & mark & entreprises & protection \\
recherche & analytica & règlement & utilisateurs \\ 
r\'eseaux & cambridge & traitement & donn\'ees \\ 
sociaux & internet & cnil & analytica \\
social & personnelles & entreprise & cambridge \\
publicit\'e & patron & droit & internet \\
annonceurs & scandale & conformit\'e & r\'eseau \\
\bottomrule
\end{tabular}
\end{table}

\subsection*{Germany} \label{resde}
Table \ref{tab:detopics} shows the results of the German model. We observe the following topics: Facebook-Cambridge Analytica scandal, explanations of GDPR for the public (i.e. what is GDPR and what does it mean for you?), GDPR jurisdiction and timeline, the narrative of government versus technology giants (e.g., Google, Facebook, and Twitter), security concerns in the technology industry, GDPR evolution (i.e., speculations about what will change in the future for businesses and society), consumer rights and protection, GDPR consequences for companies and society, motivations for GDPR, obstacles to implementation in Germany, industry relevance, and explanations from media companies about their actions regarding GDPR. The topics with the highest coherence scores include the explanations of GDPR topic, the GDPR evolution topic, the consumer rights and protection topic, the consequences of GDPR topic, and the motivations of GDPR topic. The most prevalent topics include those corresponding to GDPR evolution, consumer rights and protection, and digital security. 

\begin{table}[ht]
\centering
\caption{\label{tab:detopics}German model topic results with top 10 terms per topic and the prevalence (coherence score in parentheses) of each topic.}
\centering
\begin{tabular}{cccccc}
\toprule
Guidance & Jurisdiction & Government & Security & Evolution & Consumer Rights \\
\midrule
\textit{141 (076)} & \textit{124 (-123)} & \textit{64 (-107)} & \textit{150 (-109)} & \textit{242 (-73)} & \textit{189 (-78)} \\
\midrule
nutzer & thema & regeln & cloud & jahr & verarbeitung\\
whatsapp & datenschutzbeauftragten & gespeichert & kunden & euro & kunden \\
mail & verordnung & facebook & security & online & recht\\
facebook & uhr & vermieter & data & gut & betroffenen \\ 
kunden & mitglieder & verbraucher & sicherheit & ende & personenbezogene\\
einwilligung & ihk & adresse & management & geht & art \\ 
informationen & freitag	& recht  & anbieter & zeit & verarbeitet\\
app & sieht	& zweck	& prozent & heute & personen\\
anbieter & mitarbeiter & schutz & services & kommt & einwilligung\\
verbraucher	& umgang & eigent\"umer & software & deutschland	& mitarbeiter\\
\end{tabular}
\centering
\begin{tabular}{cccccc}
\toprule
FB/CA & Consequences & Global Motivations & Implementation & Industry & Media Explanations \\
\midrule
\textit{100 (-89)} & \textit{135 (-65)} & \textit{91 (-74)} & \textit{48 (-98)} & \textit{50 (-96)} & \textit{26 (-130)} \\
\midrule
facebook & prozent & facebook & prozent & apple & nutzer \\
usa & verordnung & zuckerberg & deutschland & iphone & personenbezogenen \\ 
zuckerberg & firmen & nutzer & grundverordnung & ger\"at  & google \\ 
europa & k\"unftig & europa & facebook & milliarden & verarbeitung\\
nutzer & euro & fragen & berg & dollar & art\\
regeln & umsetzung & beschwerden & gut & konzern & gespeichert\\ 
fragen & regeln & chef & umsetzung & nutzer & lit\\
menschen & millionen & regeln & wenig & bereits & facebook\\
nutzern	& strafen	& verbraucher	& umfrage	& millionen & rechtsgrundlage\\
google & erst & mark & russland & euro & nutzers\\
\bottomrule
\end{tabular}
\end{table}

\subsection*{UK} \label{resuk}
Table \ref{tab:uktopics} presents the results of the UK model. We observe the following topics: the Facebook-Cambridge Analytica scandal, business concerns, governance, GDPR topic areas, digital security risks (i.e. motivation for GDPR), consumer rights and protection, and GDPR legal compliance. The most coherent topics include the business, consumer rights and protection, and legal compliance topics. The legal compliance and consumer rights topics are also the most prevalent in the model. 

\begin{table}[ht]
\caption{\label{tab:uktopics}UK model topic results with top 10 terms per topic and the prevalence (coherence score in parentheses) of each topic.}
\begin{tabular}{cccccc}
\toprule
FB/CA & Government & Consumer Rights & GDPR Topics & Legal & Business \\
\midrule
\textit{23 (-31)} & \textit{21 (-32)} & \textit{87 (-36)} & \textit{23 (-10)} & \textit{120 (-6)} & \textit{51 (-18)} \\
\midrule
facebook & government & facebook & component & protection & business\\
cambridge & may & users & quote & information & year\\
analytica & brexit & privacy & emb & compani & mail\\
people & labour & information & margin & new & new\\
zuckerberg & windrush & new & chariti & personal & market\\
schroepfer & home & company & passwords & business & last\\
information & trade & people & left & may & technology\\
political & customs & also & children & breach & industry\\
company & mps & social & laws & rul & financial\\ 
wylie & people & use & system & regulation & growth\\
\bottomrule
\end{tabular}
\end{table}

\subsection*{US} \label{resus}
Table \ref{tab:ustopics} presents the results of the US model. We observe the following topics: financial implications of GDPR, Facebook-Cambridge Analytica, GDPR in the EU context, consumer rights and protection, security, legal compliance discussion, and business concerns. The most coherent topics in the US model include the EU discussion, the consumer rights and protection, and legal compliance topics. The topic prevalence analysis indicates that the legal compliance topic is also the most prevalent. Notably, this topic also appears in the UK model and is the most prevalent topic in that model as well. 

\begin{table}[ht]
\centering
\caption{\label{tab:ustopics}US model topic results with top 10 terms per topic and the prevalence (coherence score in parentheses) of each topic.}
\begin{tabular}{@{\extracolsep{5pt}} cccc}
\toprule
Finance & FB/CA & EU & Business \\
\midrule
\textit{24 (-21)} & \textit{48 (-19} & \textit{45 (-11)} & \textit{44 (20)}\\
\midrule
bank & facebook & compani & cloud \\
information	& privacy & privacy & management\\
banks & users & law & business \\
blockchain & information & european & governance \\
business & zuckerberg & new	& information \\
financial & company & protection & systems \\
credit & new & europe & enterprise \\
consumers & people & regulation & new \\
payments & compani & technology & servic \\
shar & cambridge & may & security \\
\toprule
\end{tabular}
\centering
\begin{tabular}{@{\extracolsep{5pt}} ccc}
Legal & Consumer Rights & Security\\
\midrule
\textit{70 (-6)} & \textit{47 (-16)} & \textit{46 (-19)} \\
\midrule
personal & use &security \\
information & even & percent \\
compliance & time & organizations \\
process & company & breach \\
protection & want & business \\
compani & may & risk \\
business & way & cyber \\
privacy& mobile & report \\
organizations & access & survey \\
regulation & customers & cybersecurity \\
\bottomrule
\end{tabular}
\end{table}

\subsection*{Country Comparison} 
One of the most noticeable differences across the models is the significantly wider range of topics observed in the French and German models, relative to the UK and the US ones. This could result from the size of the datasets. However, it could also result from a heightened focus of the French and German media on GDPR, its scope and myriad implications. A higher level of attention to GDPR in France and Germany might also follow from the fact that GDPR is EU legislation. Nonetheless, we suspect that this \say{EU-effect} is of secondary importance relative to the evidence that suggests these countries care more about privacy and consumer protection (or at least they care in different ways and for different reasons) than the UK and US: if we look at the most prevalent topics in each model, we see that the legal compliance topic is most prevalent in both the UK and US models. Given the results of UK and US models and knowledge of the similarity in privacy values between the two cultures, we can speculate that an over-emphasis in the media on the legal requirements of GDPR (i.e. the negative or laborious aspects of the legislation) might reflect the disinclination of UK and US cultures towards wide-sweeping privacy regulation because of the turbulence it would cause to the capitalist market interests of the two countries --- a priority we know from the literature commonly takes precedence over consumer privacy protection and state intervention on its behalf.

In contrast, the most prevalent topics in both the French and German datasets include the consumer rights and security topics. These topics include terms referring to the rights of consumers to protection and privacy, as well as the security risks associated with digital technology. Neither the French nor German model contains a topic specifically devoted to legal compliance. Instead, these models indicate higher focus on the global scope of GDPR and its broader implications for specific contexts in the French model, and friction between government and the technology industry in the German model. 

In general, the results of the French and German models indicate that the discourse on GDPR in these two countries is much more detailed and extensive than in the UK and US. Moreover, the results align with what Hofstede's scores for the four countries would imply for legislation such as GDPR. The scores predict that relative to citizens of the UK and US, citizens of France and Germany will 1) have the strongest inclination towards state-ensured privacy protection of the four countries, 2) show greater inclination towards privacy regulation, 3) prioritise privacy values over economic gains, and 4) more proactively use government regulatory powers to protect privacy. The prevalence of consumer rights and security topics in the French and German models combined with the absence of topics devoted to emphasising potential bureaucratic hang-ups, as in the UK and US models, corroborates these predictions. 

Finally, all four of the country models share the following topics: Facebook-Cambridge Analytica and consumer rights. The France, Germany, and US models all share a digital security topic, while the UK model noticeably lacks it. In order to understand more about the presence of a topic specific to the Facebook-Cambridge Analytica scandal in all four models, we read through a sample of articles with high loadings to this topic from each of the models. We found a narrative that framed GDPR as primarily a line of defence against the power of technology companies, such as Facebook and Google, instead of as the more general data protection approach that it is. GDPR is not specific to the Internet and data exchanged on its platforms; however, that is the story told by the articles we read in all of the models. Internet-related data scandals receive heavy media attention because the services involved in these scandals tend to be widely used and relied upon by a majority of the public. The fact that such scandals have become a primary focus of GDPR discourse reflects a baseline concern for social media privacy that all four countries share.

\subsection*{Temporal Analysis} \label{temporal}
\begin{figure}[ht]
\centering
\includegraphics[width=\textwidth]{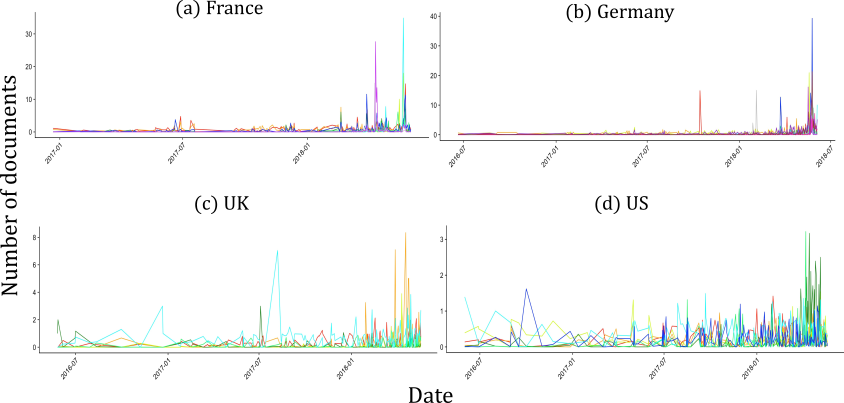}
\caption{\label{fig:temporal}Temporal evolution of the topic prevalence in each country dataset.}
\end{figure}

We count the number of documents in which that topic occurred on every day contained in the datasets. We use this metric to speculate on real-world events that might influence the patterns of emergence.
Figure \ref{fig:temporal} displays the topic emergence activity over time for each country model.

Both the French and German models exhibit low to moderate activity levels across all topics from the beginning of our collection period until the second quarter of 2018, after which we observe sharp activity increases across all topics. Nonetheless, we also observe a few spikes of abnormally high topic emergence in the German model for specific topics in isolation. That is to say, our models pick up on two different types of topic emergence: single topic surges and multiple topic surges. This suggests influences of different kinds of events that inspire the surges in media coverage attributed to specific topics. In contrast, the activity trends in the UK and US models are considerably more erratic and overall much lower than the French and German models.

In the French model, we observe two primary clusters of elevated activity (i.e. spikes in the number of documents for a given topic): 15-31 May 2018, the weeks surrounding the day on which GDPR came into effect; and April 2018, the month of Facebook CEO Mark Zuckerberg`s hearing before US Congress. The topics affiliated with these spikes of reporting include the scandal, US/UK, Facebook-Cambridge Analytica, consumer rights, and social media topics. We also observe clusters of elevated activity for specific topics in June-July 2017 (consumer rights) and early 2018 (digital security, industry, and consumer rights). The June-July 2017 surge likely results from publicity of Macron's stance on privacy regulation in the wake of his election to the presidency in May 2017. The surge in early 2018 likely results from generally renewed interest in the legislation, given that it was planned to come into effect in a few months time. 

The German model shows a majority of spikes around the GDPR effect date (25 May 2018) for the global, jurisdiction/timeline, user guidance, and security topics. However, we also see two outliers in October 2017 (the \say{consequences} topic) and February 2018 (the \say{government versus technology} topic). The spike in October 2017 coincides with the effect date of the so-called German \say{social media law} - Netzwerk Durchsetzungsgesetz (NetzDG) - aimed at fighting criminal speech on social media. The spike in the consequences topic at this point can be explained; one of the primary themes discussed in the context of NetzDG included the fines and other negative consequences that social media companies incur if they do not comply with the new legislation. This theme pervades GDPR discourse and is also the connecting theme of the other articles in the consequences topic cluster.

The spike in emergence for the government versus technology topic in early February 2018 coincides with the negotiations pertaining to Chancellor Angela Merkel's coalition government, which concluded that month. The reassertion of government regulatory power and taxation over big technology companies such as Google, Amazon, Facebook, and Apple (GAFA) was one of the main points in the negotiations between the different parties in the coalition \citep{noauthor_factbox:_2018}. Thus, both outlying spikes in the topic emergence trends appear to have correlates that help us contextualise and understand their role in the emergence dynamics of the model. 

In contrast with the French and German models, the UK and US models evidence much more scattered patterns of emergence activity. Due to the lower number of documents in these datasets, the peak activity values are also generally lower. In the UK model, the topics with the highest emergence levels include the consumer rights and legal topics. These topics surge on 7 August 2017, 29 March 2018, and 18-19 April 2018. The surge in documents within the legal topic cluster coincides directly with the publication on 7 August of the UK Statement of Intent for how to move forward on new data protection regulations in line with GDPR. The March and April 2018 surges likely result from developments in the Cambridge Analytica case. 

Neither model show spikes of activity around the GDPR effect date, which we find surprising as GDPR affects both the UK and US significantly. The highest levels of activity in the US fall in mid-late April and early-May 2018, and additionally correspond to the Facebook-Cambridge Analytica topic. These trends indicate that the Zuckerberg testimony and other Cambridge Analytica discourse appears to dominate the US GDPR media coverage.

As noted in the initial description of the UK and US results, the legal compliance topic not only emerges in both models, but is the most prevalent and among the most coherent topics in both models as well. This suggests a commonality between the UK and US GDPR media discourses, and potentially also between the privacy traditions of these nations - to the extent that they are reflected in the media. Recall that the Hofstede UAI scores for the UK and US are significantly lower than those of France and Germany, and that low UAI tends to correlate with a lower inclination towards government regulation of privacy. Furthermore, the legal compliance topic in both models includes articles that stress the complexity and burden of GDPR implementation. Thus, we suspect that more than coincidence explains the prominence of the legal compliance topic in the UK and US models: the GDPR discourse in the UK and US reflect the general lack of support for privacy regulation in these countries, as quantified by Hofstede's UAI score.

\subsection*{Topic Connectivity Analysis} \label{connect}

\begin{figure}[ht]
\centering
\includegraphics[width=\linewidth]{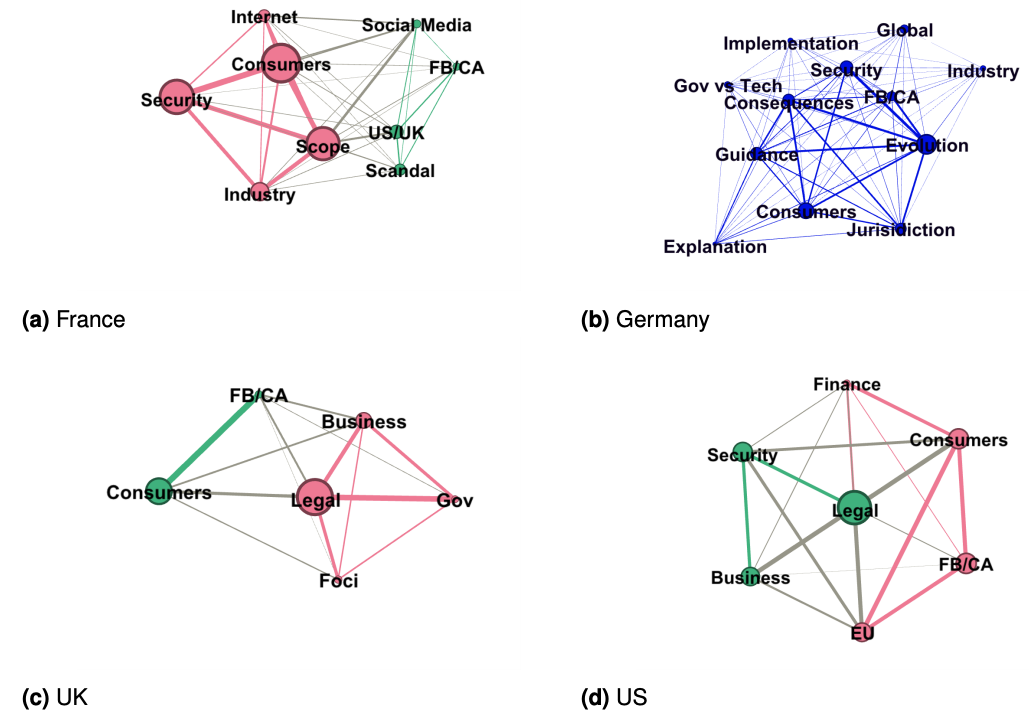}
\caption{Topic network for each country. All the topic annotations are translated to English. Different colours differentiate distinct communities detected using the Gephi implementation of the Louvain method \citep{blondel2008fast}. Node size scales in proportion with the prevalence of each topic in the model. Edge thickness similarly scales with the weight of each connection.}
\label{fig:network}
\end{figure}

In addition to analysing and comparing the emergence of topics across models, we also want to know about the connections between topics in each model. By connections we mean: how frequently do certain topics occur together in the same news item?
In order to compare the co-occurrence of topics in the documents of each corpus, we generate network visualisations of the weighted connections between the topics (see Methods). 
Figure \ref{fig:network} provides network visuals for each model.

In the French model, we identify two communities. The pink nodes, contains all of the most prevalent topics, whereas the green nodes, contains topics with very low prevalence. In our analysis of the emergence activity in the French model, we found that the Facebook-Cambridge Analytica, US/UK, and scandal topics had the highest surges of activity. These topics bear only weak connections to one another and to the nodes of the other community. These characteristics suggest that these topics are not the most central themes of the GDPR discourse in France despite their spiky popularity. The peaks of activity for these topics observed in the emergence analysis most likely represent one-off instances of French media cursorily focusing on the mostly American and British hype of the Facebook-Cambridge Analytica scandal. Meanwhile, the topics of the other community are not only more prevalent in the dataset but also more strongly connected to one another, implying that the topics of digital security, consumer rights, GDPR scope, and GDPR effects on industry comprise the central themes of French GDPR discourse. 

In contrast with the French model, the community detection algorithm run on the German model found one broad community covering all the German topics, which indicates more overlap between topics discussed in individual news items. 
Nevertheless, the topic of Facebook-Cambridge Analytica, similar to the French case, is not the main focus in the German GDPR discourse but infiltrates it nonetheless due to the global hype it caused, particularly in relation to GDPR. The primary issues in the German media in the context of GDPR appear to focus more on the practical concerns and implications of the legislation. For example, articles in the GDPR evolution topic refer to how the new legislation will affect the data economy, digital governance, and the relationship between technology companies and government entities in the years to come. Articles in the consequences topic discuss ways in which companies will have to adapt to the legislation and how GDPR will affect global commerce. 

In the UK model, we observe two topic clusters. In contrast with the French and German models, the two most prevalent topics -- legal compliance and consumer rights -- are not in the same community. The legal compliance topic has strong connections to the business, government, and GDPR foci topics in the cluster with the pink nodes. 
Intuitively, it makes sense that the consumer rights and Facebook-Cambridge Analytica topics would have a strong connection between them, indicating that the topics often co-occur in the articles within the corpus, given how heavily the consumer rights violations caused by the scandal factored into the media attention it received. It is noteworthy that the Facebook-Cambridge Analytica topic has only very weak connections to the government topic.

We also observe two distinct communities in the US topic network. The green nodes, include the legal compliance, business, and digital security topics. The pink nodes, includes the consumer rights, Facebook-Cambridge Analytica, EU, and finance topics. Similar to the UK topic network, both communities in the network contain topics of high prevalence. The US topics have the strongest interconnectivity between the two communities, with the consumer rights and EU topics sharing multiple connections with the digital security and business topics. Although the legal compliance topic has the highest prevalence in the US corpus, it does not have the most connections to other topics; rather, the consumer rights topic does. Furthermore, while the finance topic had the highest spike of emergence activity, it is the least prevalent with weakest connections in the topic network. Compared to the UK network, the US Facebook-Cambridge Analytica topic is connected to the EU topic as well as to consumer rights. These connections suggest a more diverse discourse surrounding the scandal in the US, and its relevance to GDPR, than in the UK. 
 
\section*{Conclusion} \label{conc}
As hypothesised initially, our analysis shows privacy bears different meanings and occupies different levels of sanctity across cultures. The four countries in our study each demonstrate characteristics of wealthy western civilisations but differ tremendously in their views on privacy and how they think the government should protect it. Our findings reflects these differences and indicates the issues related to GDPR and privacy protection that each country finds most pressing. Informed by Hofstede's model of cultural analysis and various comparative studies of privacy traditions, we correlate cultural features, such as uncertainty avoidance and power distance inequality, with the dynamics and emphases of each nation's GDPR discourse. 

The strongest divide exists between the nations inside and outside continental Europe: France and Germany versus the UK and the US. On a general level, France and Germany show much more media coverage related to GDPR than the UK and US. Furthermore, the range of topics present in the French and German coverage display more range, detail, and emphasis on practical implications than the UK and US counterparts. For example, while the UK and US media both seem to focus on the laborious legal aspects of the legislation, France and German media emphasised implications for consumer rights, the scope of GDPR, and how it will cause the market, the technology industry, and future legislation to evolve.
The lack of emphasis on the Facebook scandal or connections between it and other topics in both the French and German models suggests while the scandal does have privacy implications, it did not distract the media systems in these countries from the larger obstacle to privacy protection: adequate government regulation.

While the differences between these countries with respect to privacy traditions has been evident for decades if not centuries, we do not yet know how these differences will hinder or aid the acceptance and implementation of data protection regulation with global jurisdiction, such as GDPR. The results of this research provide a first step towards better understanding how different privacy traditions influence the ways in which  nations develop their stances on issues that lead to the cohesion or divergence of their respective approaches to privacy regulation in the digital age. 

Our work however has its own limitations. We cannot guarantee that the selection of articles via LexisNexis provides representative sample of each country’s media landscape. 
Moreover, topic modelling is a \say{dumb} algorithm, meaning it cannot convey the nuance of a text’s narrative or context, nor account for sentence syntax or word context. 

Despite its limitations, our selection of topic modelling  resulted in generally clear and informative topic distributions for all four of our countries. In future work of this kind, more sophisticated topic modelling algorithms that can incorporate article metadata and employ nonparametrised inference methods, such as structural topic models or stochastic block models, could increase the robustness and validity of a topic analysis of this kind of media discourse.

Furthermore, the results of our research indicate that our approach proves itself as an efficient and useful tool for real-time analysis of media discourse related to legislation. Thus, we suggest its value as a framework for incorporating the insights of media discourse into the processes of developing and implementing new legislation. Knowing exactly what the pressing, controversial, or otherwise relevant themes in a particular discourse presents tremendous value to politicians, legislators, and activists. This information would allow such actors to better understand what the public thinks and wants, thereby providing another source of public input into the legislation process.

\begin{dci}
The authors declared no potential conflicts of interest with respect to the research, authorship, and/or publication of this article.
\end{dci}

\begin{funding}
TY was partially supported by The Alan Turing Institute under the EPSRC grant [EP/N510129/1]. The Funder had no role in the conceptualization, design, data collection, analysis, decision to publish, or preparation of the manuscript.
\end{funding}

\begin{acks}
We thank Bertie Vidgen
for sharing the topic modelling code.
\end{acks}

\theendnotes

\bibliographystyle{SageH}
\bibliography{main}

\end{document}